\documentclass[mathpazo]{cicp}

\begin{document}
\title{The split-operator technique for the study of spinorial wavepacket dynamics}


\author[label1]{A. Chaves \affil{1}\comma\corrauth, G. A. Farias \affil{1}, F. M. Peeters \affil{1,2}, and R. Ferreira \affil{1,3}}
\address{\affilnum{1} Departamento de F\'isica, Universidade
Federal do Cear\'a, Caixa Postal 6030, Campus do Pici, 60455-900
Fortaleza, Cear\'a, Brazil \\
\affilnum{2} Department of Physics, University of Antwerp,
Groenenborgerlaan 171, B-2020 Antwerp, Belgium \\
\affilnum{3} Laboratoire Pierre Aigrain, Ecole Normale Superieure,
 24 Rue Lhomond, F-75005, Paris, France}

\email{{\tt andrey@fisica.ufc.br} (A.~Chaves)}


\begin{abstract}
The split-operator technique for wave packet propagation in quantum systems is expanded here to the case of propagating wave functions describing Schr\"odinger particles, namely, charge carriers in semiconductor nanostructures within the effective mass approximation, in the presence of Zeeman effect, as well as of Rashba and Dresselhaus spin-orbit interactions. We also demonstrate that simple modifications to the expanded technique allow us to calculate the time evolution of wave packets describing Dirac particles, which are relevant for the study of transport properties in graphene.
\end{abstract}

\pac{73.21.Hb, 73.63.Nm, 73.43.Cd}

\maketitle

\section{Introduction}
\label{sec1}
The time evolution of wave packets is clearly an useful tool in the study of electronic and transport properties of low dimensional systems. Investigating the propagation of wave packets in a given system allows us to obtain information about, e.g., its energy spectrum, \cite{Degani} its electric and optical conductivity, \cite{Yuan} its local density of states \cite{Fehske} and so on. In fact, wave packet dynamics methods have been successfully used in the study of the Aharonov-Bohm effect in several systems, \cite{Szafran, Szafran1, Chaves1, Romo} in the theoretical description of scanning gate microscopy experiments, \cite{Petrovic} in understanding the break of Onsager symmetry in a semiconductor quantum wire coupled to a metal, \cite{Szafran2} and in the interpretation of interference related effects in the experimentally obtained conductance of an asymmetric quantum ring, \cite{Krammer} just to mention a few examples. Lately, the interest in wave packet dynamics methods for Dirac particles has been increasing as well, \cite{Gourdeau, Demikhovskii} specially after the first experimental realization of graphene, \cite{Novoselov} a single layer of carbon atoms where low energy electrons behave as massless Dirac Fermions, thus exhibiting a series of interesting transport phenomena, such as the zitterbewegung (trembling motion) \cite{Zitterbewegung, Rusin, Zawadzki} and Klein tunneling. \cite{Katsnelson}

Several computational techniques have been developed for calculating wave packet propagation in quantum structures \cite{Peskin, Alvermann}. In fact, it is clear that, provided one has all the eigenenergies and eigenfunctions of the system, it is always possible to expand the initial wave packet in the eigenstates basis and then calculate its time evolution. However, obtaining the whole spectrum of a system is usually not an easy task, therefore, it is more convenient to look for alternative solutions to the time-dependent Schr\"odinger (or Dirac) equation. Most of the alternative techniques are based on the expansion of the time-evolution operator, in order to make it computationally easier to be applied in practical situations. Usual examples of this kind of technique are the Chebyshev polynomials expansion \cite{Talezer, Fehske2} and the split-operator technique \cite{Degani, Mclachlan, Dattoli}. The later is particularly convenient, as it splits the time evolution operator into more simple operators, written only in real or imaginary space, allowing one to avoid writing the momentum as a differential operator that has to be computationally implemented in a finite differences scheme.

In the present work, we expand the well known split-operator technique for the investigation of systems where spin-orbit interactions and Zeeman effect play an important role. Indeed, the propagation of Gaussian wave packets in a spin-orbit coupled two-dimensional electron gas (2DEG) has already been discussed in the literature, \cite{Schliemann, Biswas, Maksimova} but only with analytical methods, which, on the one hand are exact calculations but, on the other hand, they lack versatility, as they are normally too specific and problem-dependent. In the expanded split-operator technique developed here, the separation of the time-evolution operator in a series of matrices, each one only in real or reciprocal spaces, allows one to calculate the time evolution of the wave packet without using a finite differences scheme. Moreover, we demonstrate that the matrix representation of the Zeeman and spin-orbit parts of the time evolution operator comes from an \textit{exact} expansion of the exponential involved in this operator, so that the only error involved in the technique, which is proportional to the time step $\Delta t$, comes from the splitting of the exponential. Therefore, the error in the calculation is easily controlled just by setting a small value for $\Delta t$. We then apply the expanded split-operator technique to several cases, demonstrating the validity and versatility of the method in the study of the cyclotronic motion of electrons in a GaAs 2DEG under an applied magnetic field in the presence of Rashba and Zeeman coupling, as well as in the study of the zitterbewegung and Klein tunneling of wave packets in graphene.  

\section{Time Evolution Operator}
\label{sec2}

Consider an initial wave function $\Psi(\vec{r},t_0)$. By expanding this wave function in Taylor series around the initial time $t = t_0$ and defining $\Delta t = t-t_0$, one obtains
\begin{equation}
\Psi(\overrightarrow{r} ,t_0 + \Delta t) = \Psi(\overrightarrow{r}, t_0) + \sum_{n=1}^{\infty}\frac{1}{n !} \left(\frac{\partial^n \Psi}{\partial t^n}\right)_{t=t_0} \Delta t^n  .\label{eq.timeTaylor1}
\end{equation}
The Schr\"odinger equation gives $\partial \Psi / \partial t = -(i/\hbar)H \Psi$, thus,
\begin{equation}
\Psi(\overrightarrow{r} ,t_0 + \Delta t) = \sum_{n=0}^{\infty}\left[\frac{1}{n !}\left(-\frac{i}{\hbar}H \Delta t\right)^n\right] \Psi(\overrightarrow{r},t_0).  \label{eq.timeTaylor}
\end{equation}
The sum in Eq. (\ref{eq.timeTaylor}) is identified as the expansion of an exponential, from which we straightforwardly find 

\begin{equation}
\Psi(\overrightarrow{r} ,t + \Delta t) =
exp\left[-\frac{i}{\hbar}H\Delta t \right]\Psi(\overrightarrow{r}
,t). \label{eq.time}
\end{equation}

It is easy to verify that Eq. (\ref{eq.timeTaylor}) is also true for the Dirac equation, just by writing the wave function as a spinor and considering the Dirac Hamiltonian.

Our problem now consists in finding a computational technique to implement the time evolution operator into a computational routine. In some works in the literature, this is solved by considering the Cayley form for Eq. (\ref{eq.time}), which consists in an approximation of the exponential in the time evolution operator \cite{Watanabe}:
\begin{equation}
exp\left[-\frac{i}{\hbar}H\Delta t \right]\Psi(\overrightarrow{r}
,t) \simeq \frac{1+\frac{i}{2\hbar}H\Delta
t}{1-\frac{i}{2\hbar}H\Delta t}\Psi(\overrightarrow{r} ,t),
\end{equation}
so that
\begin{equation}
\left(1-\frac{i}{2\hbar}H\Delta t \right) \Psi(\overrightarrow{r}
,t + \Delta t) = \left(1+\frac{i}{2\hbar}H\Delta
t\right)\Psi(\overrightarrow{r} ,t). \label{eq.Cayley}
\end{equation}

The derivatives coming from the momentum operator in the Hamiltonian are usually written in a finite difference form, where the potential function $V(\vec{r})$ is discretized. Consequently, the wave function at each time step $t$ is also discretized in a mesh, e.g. as $\Psi(x,t) = \Psi^t_{i}$ ($i = 1, 2, ... N$) for a one-dimensional mesh with $N$ points, which can be represented as a column matrix; the multiplication in the right hand part of Eq. (\ref{eq.Cayley}) is performed, resulting in another column matrix. Equation (\ref{eq.Cayley}) is then re-written as a matrix equation, where the variables $\Psi_i^{t + \Delta t}$ in the left hand side of the equation are to be determined. By solving this matrix equation iteratively, one obtains the wave function at each time step. Difficulties appear when we try to use this method to deal with problems with more than one dimension: as we will demonstrate in details further on, the matrix equation form of Eq. (\ref{eq.Cayley}) involves a tri-diagonal matrix for a one-dimensional problem, which is very easy to handle. On the other hand, bi-dimensional problems lead to five-diagonal block matrices, whereas tri-dimensional problems involve very complicated seven-diagonal block matrices, and so on. Each of these matrices may require a lot of computational memory, or at least, may be hard and inconvenient to handle. With the split-operator method, we find a way to circumvent this difficulty, transforming an operator with any number of spatial variables into a sequence of one-dimensional operators, each one easily solved in a tri-diagonal matrix form, or even avoiding any matrix representation for the kinetic operators by using the reciprocal space through a Fourier transform of the functions.

\section{The basic split-operator technique}

First, let us separate the exponential into two parts: one of them involves only the potential energy term $V (\vec{r})$, while the other contains only the kinetic energy $T (\vec{k})$, written in the real and reciprocal spaces, respectively. The exponential of the sum $exp[A+B]$ can be separated exactly as a multiplication of exponentials $exp[A]exp[B]$ only when the operators $A$ and $B$ commute. The exponential of the Hamiltonian $H = T + V$ in the time evolution operator cannot be separated in this form, as $T$ and $V$ do not commute. Even so, we may approximate \cite{Mclachlan, Dattoli, suzuki}
\begin{equation}
e^{-\frac{i}{\hbar}H \Delta t} = e^{-\frac{i}{2\hbar}V \Delta t}e^{-\frac{i}{\hbar}T \Delta t}e^{-\frac{i}{2\hbar}H \Delta t} + O(\Delta t^3),
\end{equation}
where the terms of order larger than $\Delta t^3$ can be neglected by considering a very small time step $\Delta t$. In this way, as the terms involving the exponential of the potential are written in real space, we can simply multiply them by the wavefunction. For the kinetic energy terms, we may still use the Cayley form shown in Eq. (\ref{eq.Cayley}), however, as the kinetic energy in each direction $T_x$, $T_y$ and $T_z$, commute with each other (in the absence of magnetic fields), one can separate them exactly $exp[T] = exp[T_x+T_y+T_z] = exp[T_x]exp[T_y]exp[T_z]$. 

We start with an arbitrary wavefunction $\Psi(\overrightarrow{r},t)$, and perform the operation  \cite{Degani, Chaves1}
\begin{equation}
\Psi(\overrightarrow{r} ,t + \Delta t) =
e^{-\frac{i}{2\hbar}V\Delta t
}e^{-\frac{i}{\hbar}T\Delta t
}e^{-\frac{i}{2\hbar}V\Delta t
}\Psi(\overrightarrow{r} ,t),
\end{equation}
in order to calculate the wave function at a later time $t+\Delta t$. Discretizing the time, the potential $V$ and the wavefunction $\Psi(\overrightarrow{r},t) = |\Psi_i \rangle_t$, we first simply multiply the wavefunction by the right-hand side exponential, which involves $V$, resulting in
\begin{equation}
\xi_i = exp\left[-\frac{i}{2\hbar}V_i\Delta t \right]|\Psi_i\rangle_t.
\label{eq.SplitOpeSem1}
\end{equation}
The next step consists in multiplying $\xi_i$ by the exponential of the kinetic term. This multiplication can be performed by taking the Fourier transform of $\xi_i$, so that it is rewritten in reciprocal space, where the exponential of the kinetic part, which is also written in reciprocal space, can be simply multiplied by $\xi_i$, similarly to what we did for the potential term, as both this term and the initial wavefunction were expressed in real space. If one prefers to stay in real space, avoiding Fourier transforms, one can use the Cayley form of the exponential to obtain
\begin{equation}
\eta_i = exp\left[-\frac{i}{\hbar}T\Delta t \right]\xi_i =\left(
\frac{1+\frac{i}{2\hbar}T\Delta t}{1-\frac{i}{2\hbar}T\Delta
t}\right)\xi_i , \label{eq.kinetic1}
\end{equation}
so that
\begin{equation}
\left(1-\frac{i}{2\hbar}T\Delta t\right)\eta_i =
\left(1+\frac{i}{2\hbar}T\Delta t\right)\xi_i. \label{eq.kinetic2}
\end{equation}
As the kinetic energy, in the absence of a magnetic field, is given by
\begin{equation}
T_n = \frac{\hbar^2}{2m}\frac{d^2}{dx_n^2},
\end{equation}
where $m$ is the mass of the particle and $x_n$ is any of the spatial variables, we use the finite difference form of the derivatives, which yield an equivalent matrix equation:
\begin{eqnarray}
{\left(%
\begin{array}{ccccc}
  D_1 & D_2 & 0 & 0 & \dots \\
  D_2 & D_1 & D_2 & 0 & \dots \\
  0 & D_2 & D_1 & D_2 & \dots \\
  0 & 0 & D_2 & D_1 & \ddots \\
  0 & 0 & 0 & \ddots & \ddots \\
\end{array}%
\right)\left(%
\begin{array}{c}
  \vdots \\
  \eta_{i-1} \\
  \eta_i \\
  \eta_{i+1} \\
  \vdots \\
\end{array}%
\right)} = 
 \left(%
\begin{array}{ccccc}
   D_1' & D_2' & 0 & 0 & \dots \\
  D_2' & D_1' & D_2' & 0 & \dots \\
  0 & D_2' & D_1' & D_2' & \dots \\
  0 & 0 & D_2' & D_1' & \ddots \\
  0 & 0 & 0 & \ddots & \ddots \\
\end{array}%
\right)\left(%
\begin{array}{c}
  \vdots \\
  \xi_{i-1} \\
  \xi_i \\
  \xi_{i+1} \\
  \vdots \\
\end{array}%
\right), \label{eq.kinetic3}
\end{eqnarray}
where the matrix elements are
\begin{equation}
D_2 = -\frac{i\hbar\Delta t}{4m\Delta x_n^2} \quad\quad\quad\quad D_1
= 1+\frac{i\hbar\Delta t}{2m\Delta x_n^2}
\end{equation}
and
\begin{equation}
D_2' = \frac{i\hbar\Delta t}{4m\Delta x_n^2} \quad\quad\quad\quad D_1'
= 1-\frac{i\hbar\Delta t}{2m\Delta x_n^2},
\end{equation}
with $\Delta x_n$ step in the $x_n$-direction. The right-hand side of Eq. (\ref{eq.kinetic3}) can be directly multiplied, because we already know $\xi_i$ for any $i$ of the grid. The remaining tridiagonal matrix equation for $\eta_i$ must then be solved numerically, which can be easily done by using existing computational routines.\cite{NR}

By solving this matrix equation, we obtain $\eta_i$, which will be used to finally calculate the wave function in $t + \Delta t$, by direct multiplication
\begin{equation}
|\Psi_i\rangle_{t + \Delta t} = exp\left[-\frac{i}{2\hbar}V_i\Delta t
\right] \eta_i.
\end{equation}
As previously mentioned, if the problem requires more spatial variables, one can repeat the procedure in Eqs. (\ref{eq.kinetic1}) - (\ref{eq.kinetic3}) for the kinetic energy in each direction. In this way, one can solve problems with any number of spatial variables just by performing calculations with tridiagonal matrices, one for each dimension, instead of performing operations with giant matrices, involving the discretization of all coordinates at once, which is usually done when applying the Cayley form without the split-operator technique. Alternatively, one can perform a Fourier transform and apply the exponentials of the kinetic energy operator in each direction without the use of any matrix equation.

\section{Spin dependent Hamiltonians}

There is a special class of Hamiltonians that can be treated in a
very simple way with the split-operator technique, namely, Hamiltonians that can be written in terms of Pauli matrices
\begin{equation}
\overrightarrow{\sigma} = \sigma_x \hat{i} + \sigma_y \hat{j} +
\sigma_z \hat{k},
\end{equation}
where
\begin{equation}
\sigma_x = \left(
\begin{array}{cc}
  0 & 1 \\
  1 & 0 \\
\end{array}
\right),  \sigma_y = \left(\begin{array}{cc}
  0 & -i \\
  i & 0 \\
\end{array}
\right),  \sigma_z = \left(\begin{array}{cc}
  1 & 0 \\
  0 & -1 \\
\end{array}
\right).
\end{equation}

Several Hamiltonians can be written in this form, \emph{e. g.} the
Hamiltonian for the Zeeman effect, $H_Z = 0.5 g\mu \overrightarrow{B} \cdot \overrightarrow{\sigma}$, and those describing spin-orbit coupling, such as the
Dresselhaus $H_D = 0.5 \alpha_D \overrightarrow{\Omega}(p) \cdot \overrightarrow{\sigma}$ and Rashba $H_F = 0.5 \alpha_R \overrightarrow{p} \times \overrightarrow{\sigma}$ Hamiltonians.
Besides, the Hamiltonian describing graphene in the continuum
model can also be written in this way \cite{CastroNetoReview}: $H = v_F \hbar
\overrightarrow{k} \cdot \overrightarrow{\sigma} +
F(\vec r)v_f^2\sigma_z$, where $F(\vec r)$ is a space-dependent mass term. \cite{Zhou}

Let us assume that the Hamiltonian can be written as
\begin{equation}
H = \overrightarrow{W} \cdot \overrightarrow{\sigma},
\end{equation}
the time evolution operator is then
\begin{equation}
\exp\left[-\frac{i}{\hbar}H\Delta t\right] =
\exp\left[-\frac{i}{\hbar} \Delta t \overrightarrow{W} \cdot
\overrightarrow{\sigma}\right] = \exp\left[-i\overrightarrow{S}
\cdot \overrightarrow{\sigma}\right]. \label{eq.opertimespin}
\end{equation}
Writing the time evolution operator in this form, it is
straightforward to see that the expansion of this exponential is
\begin{eqnarray}
\exp\left[-i\overrightarrow{S} \cdot \overrightarrow{\sigma}\right]
=
\sum_{n=0}^{\infty}\frac{(-i\overrightarrow{S}\cdot\overrightarrow{\sigma})^n}{n!} = 
\sum_{k=0}^{\infty}\frac{(-1)^k(\overrightarrow{S}\cdot\overrightarrow{\sigma})^{2k}}{(2k)!}
- i
\sum_{k=0}^{\infty}\frac{(-1)^k(\overrightarrow{S}\cdot\overrightarrow{\sigma})^{2k+1}}{(2k+1)!}.
\label{eq.expoperspin}
\end{eqnarray}
Now, we take advantage of two well known properties of the Pauli
matrices: $\sigma_i \sigma_i = I $ and $[\sigma_i, \sigma_j]_+ =
0$, where $I$ is the identity matrix, to obtain
\begin{equation}
(\overrightarrow{S} \cdot \overrightarrow{\sigma})^{2k} = S^{2k}I,
\quad (\overrightarrow{S} \cdot
\overrightarrow{\sigma})^{2k+1} = S^{2k}(\overrightarrow{S} \cdot
\overrightarrow{\sigma}) \label{eq.propsigmas}
\end{equation}

With these properties, the expansion of the time evolution operator in Eq.
(\ref{eq.expoperspin}) can be re-written as
\begin{eqnarray}
\exp\left[-i\overrightarrow{S} \cdot \overrightarrow{\sigma}\right]
= \left(
\begin{array}{cc}
  \cos(S) & 0 \\
  0 & \cos(S) \\
\end{array}
\right) 
-i\frac{\sin(S)}{S}\left(
\begin{array}{cc}
  S_z & S_x - iS_y \\
  S_x + iS_y & -S_z \\
\end{array}
\right) = M, \label{eq.opertimespinFin}
\end{eqnarray}
where $S$ is the modulus of the
vector $\overrightarrow{S}$ defined in Eq.
(\ref{eq.opertimespin}) and $S_i$ ($i = x, y, z$) is its component in the $i$-direction. Thus, the time evolution
operation is represented by a simple matrix multiplication. Notice
that this matrix form is an exact representation of the time evolution
operator, including all the terms of the expansion of the
exponential.

In the cases where $\overrightarrow{S}$ depends on
the wave vector $\overrightarrow{k}$, one may have problems in
finding the modulus $S$, as well as in calculating its sine and
cosine, if one deals with the operators in real space,
where the components of $\overrightarrow{k}$ are written as
spatial derivatives. Hence, in these cases, one should stay in momentum space, by performing a Fourier transform on the wavefunctions, in order to rewrite them in a space where the $k_i$ are numbers, instead of derivatives, so that one needs only to multiply the wavefunctions by the matrix elements of Eq. (\ref{eq.opertimespinFin}).

\section{Imaginary time evolution: obtaining eigenstates}

The eigenstates of a given Hamiltonian can be obtained through the split-operator technique described above. In order to obtain the ground state, one must simply propagate an arbitrary wave function in the imaginary time domain - since the eigenstates form a complete orthonormal basis, any arbitrary wave function can be written as a linear combination of these eigenstates:
\begin{eqnarray}
|\Psi\rangle_t = \sum_{n=0}^{\infty} a_n e^{-\frac{iE_n
t}{\hbar}}|\Phi_n> \label{eq.sumeigen}
\end{eqnarray}
where $\Phi_n$ and $E_n$ are, respectively, the eigenfunctions and eigenenergies of the $n$-th eigenstate. By defining $\tau = it$, 
\begin{eqnarray}
|\Psi\rangle_t = \sum_{n=0}^{\infty} a_n e^{-\frac{E_n
 \tau}{\hbar}}|\Phi_n\rangle =  e^{-\frac{E_0 \tau}{\hbar}} \left[ a_0|\Phi_0> + \sum_{n=1}^{\infty} a_n
 e^{-\frac{(E_n - E_0)
 \tau}{\hbar}}|\Phi_n\rangle\right]  \label{eq.sumtau}
\end{eqnarray}
so that, when $\tau \rightarrow \infty$, the ground state term in the sum becomes dominant over the other terms, since $E_n - E_0 > 0$ for $n > 0$. Therefore, starting the propagation with any initial wave function, this function must converge to the eigenstate as imaginary time $\tau$ elapses. Excited states can then be obtained by means of a Gram-Schmidt orthonormalization: if a given initial wave function is orthonormal to the ground state, for example, the ground state wave function cannot be present in the linear combination that describes such initial function in Eq. (\ref{eq.sumeigen}), thus, the lowest energy term in this sum has energy $E_1$ and, consequently, the wave function must converge to $|\Phi_1\rangle$ as $\tau \rightarrow
\infty$. In order to obtain $|\Phi_2\rangle$, one starts with a wave function that is orthonormal to both $|\Phi_1\rangle$ and $|\Phi_0\rangle$, and so on.

Notice that this method would not be suitable for calculating eigenstates in graphene. This is due to the fact that the Hamiltonian in graphene has an energy spectrum that covers both negative and positive energies, whereas the eigenstates of interest are normally around the Fermi level $E_F = 0$. The procedure described above would lead to the lowest energy state, which, in the continuum model of graphene, is $E \rightarrow -\infty$, deep in the ''Dirac sea'' and far from $E_F$. 

\section{Examples of application}

In what follows, we investigate the time evolution of a Gaussian-like spinor multiplied by a plane wave in the $y$-direction
\begin{equation}\label{eq:initial}
\Psi(x,y,0) = N\left(\begin{array}{c}
\phi_A\\
\phi_B
\end{array}\right)\exp\left(ik_0y - \frac{x^2}{2 d_x^2}- \frac{y^2}{2 d_y^2}\right),
\end{equation}
where $N$ is a normalizing factor and $k_0 = \sqrt{2mE/\hbar}$ is its wave number, in two systems of current interest: a semiconductor quantum dot with Zeeman and spin-orbit interactions and a monolayer graphene with potential barriers.

\subsection{Zeeman and Spin-orbit interactions in quantum dots revisited}

The spin-dependent split-operator formalism can be easily applied for
studying e.g. a planar quantum dot in the
presence of an applied magnetic field. The confinement potential we
consider consists of a step in the radial direction, so that
$V(x,y) = 0$ for $R^2 > x^2+y^2$ and $V(x,y) = V_e$ otherwise,
where $R$ is the dot radius and $V_e$ is the conduction
band-offset. We consider here a circular 2D dot for simplicity, but an arbitrary dot geometry can be considered as well, just with straightforward adaptation.

Let us define $H_0$ as the Hamiltonian for the electrons
confinement in such a quantum dot, considering a magnetic field
$\overrightarrow{B} = B \hat{z}$, described by a vector potential
in the symmetric gauge, $\overrightarrow{A} = (-By/2, Bx/2, 0)$,
in the absence of Zeeman and spin-orbit interactions. The Hamiltonian
for the Zeeman effect is given by
\begin{equation}
H_Z = \frac{1}{2}g\mu B \sigma_z,
\end{equation}
where $g$ is the effective Land\'e factor, whereas the quadratic Dresselhaus Hamiltonian is
\begin{equation}
H_D = \alpha_D \left[-\left(p_x + \frac{eBy}{2}\right)\sigma_x +
\left(p_y - \frac{eBx}{2}\right)\sigma_y\right],
\end{equation}
and the Rashba Hamiltonian can be re-written as
\begin{equation}
H_R = \alpha_R \left[\left(p_y - \frac{eBx}{2}\right)\sigma_x -
\left(p_x + \frac{eBy}{2}\right)\sigma_y\right],
\end{equation}
so that the Hamiltonian for this system is given by $H = H_0 + H_Z
+ H_D + H_R$. Each interaction can be effectively 'turned off' in the calculations just by setting its coefficient ($g$, $\alpha_D$ and $\alpha_R$, respectively) to zero.

We now use the split-operator technique developed in the previous Sections for separating the Zeeman and spin-orbit terms from $H_0$, so that the time evolution operator is approximated as
\begin{equation}
e^{-\frac{i\Delta t}{\hbar}H} =
e^{-\frac{i\Delta
t}{2\hbar}(H_D+H_R+H_Z)}e^{-\frac{i\Delta
t}{\hbar}H_0} e^{-\frac{i\Delta
t}{2\hbar}(H_D+H_R+H_Z)}.
\end{equation}
The $H_{S} = H_D+H_R+H_Z$ part can be re-written as
\begin{equation}
H_S = (\alpha_Rp_y - \alpha_Dp_x)\sigma_x + (\alpha_Dp_y -
\alpha_Rp_x)\sigma_y 
- \frac{eB}{2}(\alpha_Rx+\alpha_Dy)\sigma_x -
\frac{eB}{2}(\alpha_Ry+\alpha_Dx)\sigma_y + \frac{1}{2}g\mu B
\sigma_z.
\end{equation}
so that 
\begin{equation}
\exp{\left[-\frac{i\Delta t}{2\hbar}(H_D+H_R+H_Z)\right]} =
\exp[-i\left(\overrightarrow{v_1} \cdot \overrightarrow{\sigma}+
\overrightarrow{v_2} \cdot \overrightarrow{\sigma}\right)],
\end{equation}
where the vectors $\overrightarrow{v_1}$ and $\overrightarrow{v_2}$ contain only terms in real and reciprocal space, respectively:
\begin{equation}
\overrightarrow{v_1} = \frac{\Delta
t}{2\hbar}\left[(\alpha_Rp_y-\alpha_Dp_x),
(\alpha_Dp_y-\alpha_Rp_x), 0\right] 
\overrightarrow{v_2} = \frac{\Delta
t}{2\hbar}\left[-\frac{eB}{2}(\alpha_Rx+\alpha_Dy),
-\frac{eB}{2}(\alpha_Ry+\alpha_Dx), \frac{1}{2}g\mu B\right].
\end{equation}
Due to the non-commutativity between the terms
$[\overrightarrow{v_1}\cdot\overrightarrow{\sigma},
\overrightarrow{v_2}\cdot\overrightarrow{\sigma}] \neq 0$, the split-operator technique must be used once more, now for separating the exponentials for each of these vectors, consequently, completely separating the exponentials containing terms in real and reciprocal space
\begin{eqnarray}
\exp{\left[-\frac{i\Delta t}{2\hbar}(H_D+H_R+H_Z)\right]} = \exp\left[-\frac{i}{2}\overrightarrow{v_2} \cdot
\overrightarrow{\sigma}\right]\exp\left[-i\overrightarrow{v_1}
\cdot
\overrightarrow{\sigma}\right]\exp\left[-\frac{i}{2}\overrightarrow{v_2}
\cdot \overrightarrow{\sigma}\right].
\end{eqnarray}
Finally, this procedure leads to a form of the time evolution operator that is compatible with Eq. (\ref{eq.opertimespinFin}). The time evolution of a spinor-like wave packet $|\Psi \rangle_{t_0} = (u ~~ d)^T f(\vec{r},t_0)$, where $u$ ($d$) stands for the upper (lower) component of the spinor, is therefore calculated as
\begin{eqnarray}
|\Psi \rangle_{t+\Delta t} =
e^{-\frac{i}{2}\overrightarrow{v_2}\cdot\overrightarrow{\sigma}}e^{-i\overrightarrow{v_1}\cdot\overrightarrow{\sigma}}e^{-\frac{i}{2}\overrightarrow{v_2}\cdot\overrightarrow{\sigma}}
e^{-\frac{i\Delta t}{\hbar}H_0} 
e^{-\frac{i}{2}\overrightarrow{v_2}\cdot\overrightarrow{\sigma}}e^{-i\overrightarrow{v_1}\cdot\overrightarrow{\sigma}}e^{-\frac{i}{2}\overrightarrow{v_2}\cdot\overrightarrow{\sigma}}|\Psi
\rangle_t
\end{eqnarray}
where the exponentials involving $\vec{v_1}$ and $\vec{v_2}$ can be re-written as matrices, according to Eq.
(\ref{eq.opertimespinFin}),
\begin{equation}
|\Psi \rangle_{t+\Delta t} = M_2 \cdot M_1 \cdot M_2 \cdot
e^{-\frac{i\Delta t}{\hbar}H_0} \cdot M_2 \cdot M_1 \cdot M_2
|\Psi\rangle_t , \label{eq.dotspinFin}
\end{equation}
where $M_1$ and $M_2$ are the matrix representations of $\exp\left[-i\vec{v_1}\cdot\vec{\sigma}\right]$ and $\exp\left[-i\vec{v_2}\cdot\vec{\sigma}/2\right]$, respectively.

Thus, the time evolution of a spinor-like wave packet in the presence of Zeeman and spin-orbit effects is (exactly) obtained simply by the sequence of matrix multiplications in Eq. (\ref{eq.dotspinFin}). Notice that a (inverse) Fourier transform must be taken before (after) operating with the $M_1$ matrix, since this matrix contains only terms in reciprocal space, whereas $M_2$ is written in real space.

\begin{figure}[!bpht]
\centerline{\includegraphics[width=0.7\textwidth]{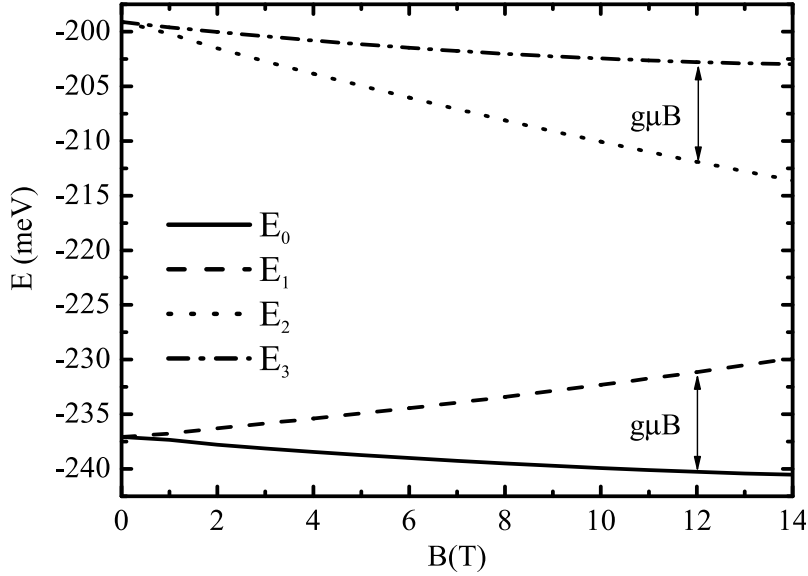}}
\caption{Energy levels of a planar quantum dot with radius $R$ = 100 \AA\,, as a function of the magnetic field intensity, in the absence of spin-orbit coupling.}
\label{fig:zeemandot}
\end{figure}

As a test case, let us study only the Zeeman effect (i.e., $\alpha_R = \alpha_D = 0$) in the eigenstates of an electron confined in a planar GaAs circular dot with radius $R$ = 100 \AA\,, in the presence of a perpendicular magnetic field applied in the $z$-direction. The four low-lying energy levels of this system, numerically obtained by the evolution of four orthogonal arbitrary wave packets in imaginary time, as discussed in the previous section, are shown in Fig. \ref{fig:zeemandot}. Notice the magnetic field dependent separation between the pairs of states $E_1 - E_2$ and $E_3 - E_4$, such that $\Delta E_{n-m} = g\mu B$, just as expected for the Zeeman effect.

We now investigate the real time dependence of the $z$-component of the electron spin in such a system. If we keep the magnetic field applied in $z$-direction, as before, the $|+\rangle = (1 ~~ 0)^T$ and $|-\rangle = (0 ~~ 1)^T$ spinor states, with $\langle \sigma_z \rangle$ = 1 and -1, respectively, are the eigenstates of the system, so that the $z$-component of the electron spin will remain the same along the whole time evolution. However, considering an in-plane magnetic field, e.g. in the $x$-direction, these states are not eigenstates of the system, since the Zeeman Hamiltonian in this case is $H_{Z} = g \mu B \sigma_x/2$, so that in the basis of the $z$-component spinors $|+\rangle$ and $|-\rangle$, the eigenstates of $\sigma_x$ are
\begin{equation}
|1\rangle = \frac{1}{\sqrt{2}}(|+\rangle + |-\rangle), \quad \quad |2\rangle =
\frac{1}{\sqrt{2}}(|+\rangle - |-\rangle).
\end{equation}
Therefore, a purely up or purely down state would be written in the basis of the $\sigma_x$ eigenstates as
\begin{eqnarray}
|+\rangle = \frac{1}{\sqrt{2}}(e^{-iE_1t/\hbar}|1\rangle +
e^{-iE_2t/\hbar}|2\rangle), \nonumber \\
|-\rangle = \frac{1}{\sqrt{2}}(e^{-iE_1t/\hbar}|1\rangle - e^{-iE_2t/\hbar}|2\rangle)
\label{eq.eigensigmax0},
\end{eqnarray}
respectively, which can be easily re-written as
\begin{equation}
|\pm\rangle = \frac{1}{\sqrt{2}}(|1\rangle \pm e^{-i(E_2-E_1)t/\hbar}|2\rangle).
\label{eq.eigensigmax}
\end{equation}

\begin{figure}[!bpht]
\centerline{\includegraphics[width=0.7\textwidth]{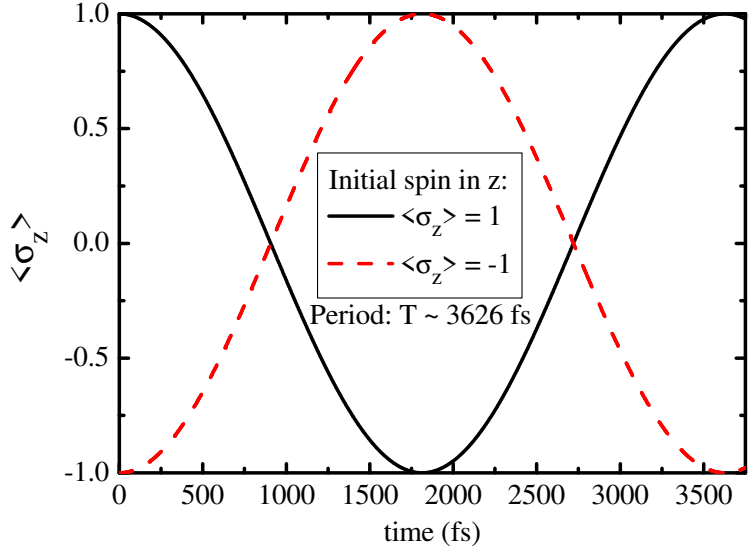}}
\caption{$z$-component of the electron spin as a function of time in a quantum dot under an applied magnetic field parallel to the quantum dot plane, considering only the Zeeman effect, for initial functions in the spin up (black) and down (red) states.}
\label{fig:spinsigmax}
\end{figure}

The numerically obtained time evolution of the $z$-component of the spin in such a system with magnetic field $B$ = 1 T applied parallel to the quantum dot plane is shown in Fig. \ref{fig:spinsigmax}, for initial wave functions describing electrons with spin states such that $\langle \sigma_z \rangle$ = 1 (black, spin up) or -1
(red, spin down). We observe an oscillatory behavior with period $T =$
3626 fs for this spin component. Notice that the exponential in Eq. (\ref{eq.eigensigmax}) is indeed a term that periodically oscillates between -1 and +1, leading to a sum or difference between $|1\rangle$ and $|2\rangle$, which, if compared to Eq. (\ref{eq.eigensigmax0}), are easily identified as the up or down spin states, respectively. Actually, this oscillation is closely related to the spin precession that is observed when the electron spin does not point towards the magnetic field direction. In summary, this analysis of Eq. (\ref{eq.eigensigmax}) explains the periodic oscillations in the time evolution of the $z$-component of the electron spin in Fig. \ref{fig:spinsigmax}. Moreover, the analysis of Eq. (\ref{eq.eigensigmax}) demonstrates an oscillation period given by $T = 2\pi\hbar/(E_2-E_1)$; by substituting $E_2 - E_1 = g\mu B$, one obtains $T = 3626$ fs, the same period numerically found in Fig. \ref{fig:spinsigmax} by means of the split-operator technique. Such a good agreement between analytical and numerical results helps to validate the extension of the split-operator technique for spin-dependent Hamiltonians developed in this paper.

\begin{figure}[!bpht]
\centerline{\includegraphics[width=0.7\textwidth]{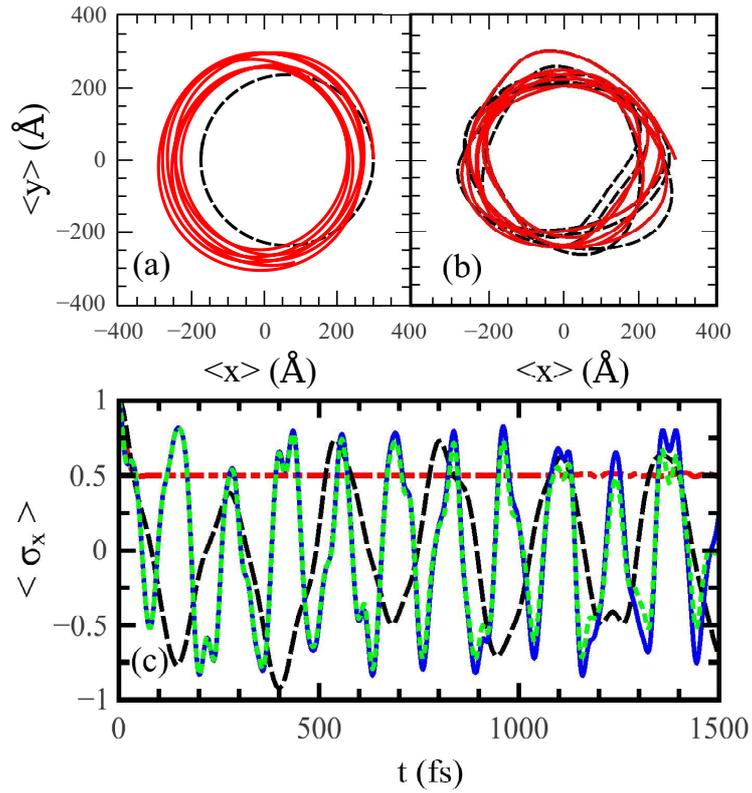}}
\caption{(a) Trajectories of a $E = 10$ meV, $d_x = d_y = 100$ \AA\, Gaussian wave packet moving in a GaAs 2DEG, under a $B = 10$ T magnetic field, in the absence of spin-orbit and Zeeman effects (black, dashed) and in the presence of a Rashba spin-orbit term with $\alpha_R$ = 1 eV\AA\,(red, solid). (b) The same as (a), but also with a Dresselhaus spin-orbit term $\alpha_D$ = 2 eV\AA\,, for $g = 0$ (black, dashed) and -0.044 (red, solid). (c) Expectation value of the $z$-component of spin as a function of time for $\alpha_R = 1$ eV\AA\,, considering $g = 0$ and $\alpha_D$ = 0.5  (black, dashed), 1 (red, dashed-dotted) and 2 eV\AA\, (blue, solid), and for $g = -0.044$ and $\alpha_D = 2$ eV\AA\, (green, dotted).}
\label{fig:SO}
\end{figure}

Let us now investigate the case where spin-orbit effects are present. Figures \ref{fig:SO}(a) and (b) show the trajectories ($\langle x \rangle$ and $\langle y \rangle$ as time elapses) performed by the cyclotron motion of an electron, described by the Gaussian-like wave packet of Eq. (\ref{eq:initial}) with $\Phi_A = \Phi_B = 1$ (i.e. $\langle \sigma_x\rangle = 1 $), $d_x = d_y = 100$ \AA\,, and $E = 10$ meV, in a GaAs 2DEG, in the presence of a $B = 10$ T perpendicularly applied magnetic field. The result in the absence of any spin-orbit or Zeeman effects is shown by the dashed black line in Fig. \ref{fig:SO}(a), which is simply a circular orbit. In the presence of Rashba spin-orbit effect with $\alpha_R$ = 1 eV\AA\,, distortions of the circular trajectory are observed, which is similar to what was obtained in Ref. \cite{Schliemann}. The distortions are stronger in the presence of both Rashba and Dresselhaus spin-orbit terms, as shown in Fig. \ref{fig:SO}(b), where the trajectories for $\alpha_R$ = 1 eV\AA\, and $\alpha_D =$ 2 eV\AA\, are shown by the black dashed and red solid lines, for Zeeman terms with $g = 0$ and -0.044, respectively. The presence of the Zeeman effect modifies the electron spin dependence on time, as shown by the blue solid and green dotted lines in Fig. \ref{fig:SO}(c), which correspond to the situations represented by black dashed (with $g = 0$) and red solid (with $g$ = -0.044) curves in Fig. \ref{fig:SO}(b), respectively. Although the changes observed in $\langle \sigma_x \rangle$ are small, they still significantly modify the trajectory, since this trajectory strongly depends on the spin-orbit terms, closely related to $\langle \sigma_x \rangle$. In Ref. \cite{Schliemann}, it was mentioned that exact analytical solutions for such a system in the presence of both Rashba and Dresselhaus spin-orbit coupling terms are not possible, and that a particular situation would be reached if both terms had the same magnitude, i.e. if $\alpha_R = \alpha_D$, due to the existence of a new conserved spin operator in this case.  \cite{Badalyan1, Badalyan2} Our numerical approach allows us to investigate this situation, which is also shown in Fig. \ref{fig:SO}(c), where we fix $\alpha_R$ = 1 eV\AA\, and consider three values for the Dresselhaus term: $\alpha_D = 0.5$ (black, dashed), 1 (red dashed-dotted) and 2 eV\AA\, (blue, solid). A very interesting result is observed for $\alpha_D = \alpha_R = 1$ eV\AA\,: $\langle \sigma_z \rangle$ rapidly reaches 0.5 and does not change in time afterwards, which may be a consequence of the new conserved spin operator mentioned in Ref. \cite{Schliemann}.

\subsection{Dirac Hamiltonian for graphene}

Low energy electrons in monolayer graphene behave as massless Dirac fermions with Fermi velocity $v_F = 3t/a\hbar$. \cite{CastroNetoReview} The Dirac Hamiltonian $H_D$ for graphene can be separated as $H_D = H_k + H_r$, where $H_k = \hbar v_F\vec{\sigma}\cdot\vec{k}$ keeps only the terms depending on the wave vector $\vec{k}$, whereas $H_r = v_Fe\vec{\sigma}\cdot\vec{A} + V\textbf{I} + F(\vec r)\sigma_z$ depends on the real space coordinates $x$ and $y$, where $\vec{A}$ is the vector potential. Using the split-operator technique for the time evolution of a wave packet in graphene, one obtains, approximately,
\begin{eqnarray}
\exp\left[-\frac{i\Delta t}{\hbar}\left(H_k + H_r\right)\right]
\approx \nonumber \quad\quad\quad\quad
\\
\exp\left[-\frac{i \Delta t}{2\hbar}H_r\right]\exp\left[-\frac{i
\Delta t}{\hbar}H_k\right]\exp\left[-\frac{i \Delta
t}{2\hbar}H_r\right]
\end{eqnarray}
Using Eq.(\ref{eq.opertimespinFin}), these exponentials are re-written, respectively, as \cite{Rakhimov}

\begin{equation}
\mathcal M_r = \left[\cos\left(\varrho \right)\textbf{I} -i \frac{\sin\left(\varrho \right)}{\varrho}\left(%
\begin{array}{cc}
   {\textsf{M}} & {\textsf{A}_x-i\textsf{A}_y} \\
 {\textsf{A}_x+i\textsf{A}_y} & -{\textsf{M}} \\
\end{array}%
\right)\right]e^{-\frac{i\Delta t}{2\hbar}V},
\end{equation}
\begin{equation}
\mathcal{M}_k = \cos(\kappa)\textbf{I} -i\frac{\sin(\kappa)}{\kappa}\left(%
\begin{array}{cc}
  0 & \kappa_x-i\kappa_y \\
  \kappa_x+i\kappa_y & 0 \\
\end{array}%
\right),
\end{equation}
where $\vec{\kappa} = \Delta t v_F\vec{k}$, $\kappa =
|\vec{\kappa}|=\Delta t
v_F\sqrt{k_x^2+k_y^2}$, $\vec \varrho = (\textsf{A}_x, \textsf{A}_y, \textsf{M})$, $\varrho = |\vec \varrho|$ and we define the dimensionless quantities $\vec{\textsf{A}} =\Delta t v_F e\vec A \big/2\hbar$ and $\textsf{M} = \Delta t M \big/2\hbar$. Therefore, the time evolution of a wave packet $\Psi_D(x,y) = (\phi_A ~~ \phi_B)^T\Psi(x,y)$ can be calculated through a series of matrix multiplications:
\begin{equation}
\Psi(\vec{r},t+\Delta t) =
\mathcal{M}_r\cdot\mathcal{M}_k\cdot\mathcal{M}_r\Psi(\vec{r},t)
+ O(\Delta t^3).
\end{equation}
The multiplications with $\mathcal{M}_k$ are performed in reciprocal space, i.e. taking a Fourier transform of the functions involved. In the absence of a magnetic field, mass and external potentials, one has $\mathcal{M}_r$ = \textbf{I} and, consequently,
\begin{equation}\label{DiracExact}
\Psi(\vec{r},t+\Delta t)
=\mathcal{M}_k\Psi(\vec{r},t),
\end{equation}
where the matrix multiplication in reciprocal space leads to an exact time evolution for the wave packet, since there is no error induced by non-commutativity between the operators in this case. Thus, within the split-operator technique, in the presence of an external potential and/or magnetic field, one can control the accuracy of the results by adjusting $\Delta t$, whereas in the absence of fields and/or potentials, the problem is exactly solved by a simple matrix multiplication, for any value of $\Delta t$.

\begin{figure}[!bpht]
\centerline{\includegraphics[width=0.7\textwidth]{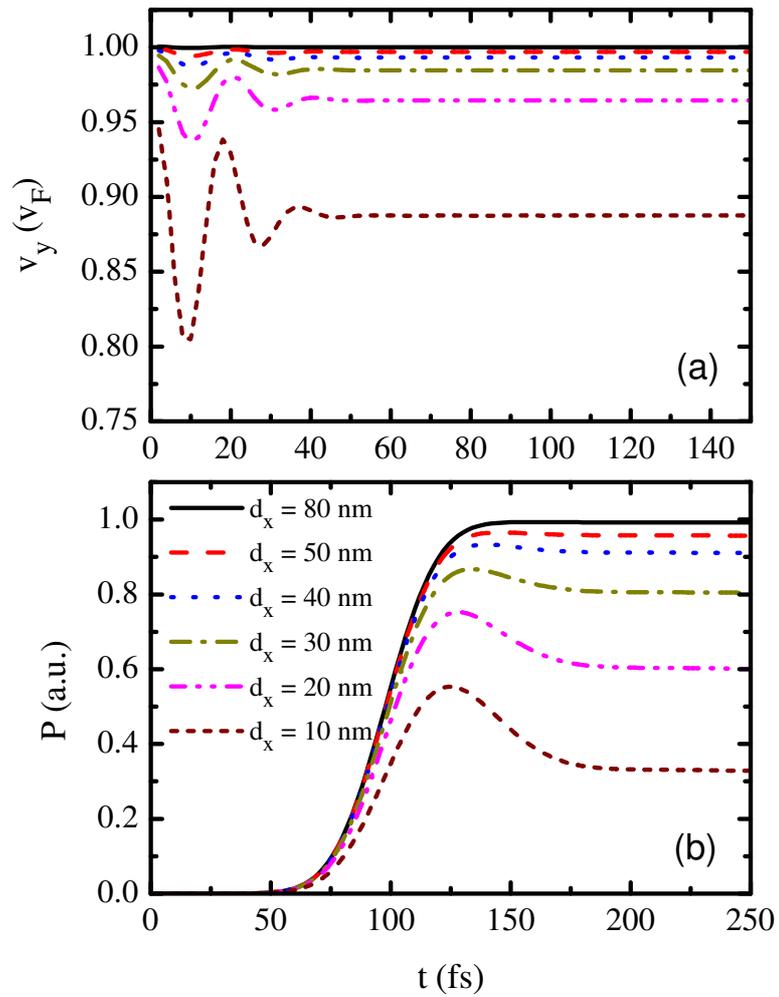}}
\caption{(a) Propagation velocity oscillations (zitterbewegung) in the absence of external potentials and (b) probability of finding the electron inside a potential barrier (Klein tunnelling), as a function of time, for $E = 100$ meV, $d_y = 20$ nm and different values of wave packet width in the $x-$direction.}
\label{fig:graphene}
\end{figure}

As an example of a practical application of the split-operator technique for graphene, let us investigate two well known effects observed for Dirac particles and, consequently, for low-energy electrons in graphene: (i) the zitterbewegung, i.e. a natural trembling motion of the wave packet \cite{Zawadzki, Frolova, Zitterbewegung}, and (ii) the Klein tunnelling \cite{Allain, Beenakker, PereiraReview}. The former manifests itself as a deformation of the wave packet and as oscillations on the average position and group velocity. Notice that the dispersion of a wave packet comes from the linear dependence of the group velocity on the momentum, which is present for Schr\"odinger particles, but not for massless Dirac particles. Thus, there should be no dispersion, i.e. no distortion on the wave packet as time elapses for graphene, which means that any distortion observed in this case comes only from zitterbewegung. Within the Heisenberg picture, the propagation velocity in the $y$-direction is obtained as 
\begin{equation}
v_y = \frac{d y}{dt} = \frac{1}{i\hbar}\left[y,H\right] = v_F\sigma_y.
\end{equation}
For simplicity, let us consider the most common case of zero mass $F(\vec{r}) = 0$. The time-dependence of this velocity is given by
\begin{equation}
\frac{d \sigma_y}{dt} = \frac{1}{i\hbar}\left[\sigma_y,H\right] = v_F k_x\sigma_z.
\end{equation}
If the wave packet contains non-zero $k_x$ components, $\langle \sigma_y \rangle$ will not be a constant of motion and, consequently, $v_y$ will vary in time. This is demonstrated in Fig. \ref{fig:graphene}(a), where we observe oscillations in the propagation velocity of a $E = 100$ meV, $d_y = 20$ nm, $\phi_A = 1$ and $\phi_B = i$ wave packet, as the time evolves. These oscillations are much stronger when the wave packet is narrower in the $x$-direction, i.e. when $d_x$ is smaller, since in this case the wave packet is represented by a large distribution of $k_x$ in reciprocal space.

Figure \ref{fig:graphene}(b) shows the probability $P = \int_{-\infty}^{\infty}dx\int_{0}^{\infty}dy|\Psi|^2$ of finding the propagating electron represented by the wave function in Eq. (\ref{eq:initial}), with the same parameters as in (a), inside a step barrier region of height $V_0 = E = 100$ meV at $y > 0$. The theory of Klein tunnelling states that an electron with normal incidence on such a barrier is perfectly transmitted, leading to $P \rightarrow 1$. Our results demonstrate that this is only true if the electron is represented by a plane wave, or by a wave front, i.e. with $d_x \rightarrow \infty$, so that its wave function contains a single value of momentum in the $x$-direction, namely $k_x = 0$. As $d_x$ decreases, the wave packet becomes wider in the $k_x$ direction in reciprocal space, leading to parts of the wave packet that does not effectively reach the barrier with normal incidence (namely, with $k_x \neq 0$), which reduces the transmission probability. This effect is much stronger for $V_0 = E$, since the dependence of the Klein tunneling probability on the incidence angle becomes negligible for an electron energy $E$ far from the barrier height $V_0$. \cite{PereiraReview, Matulis}

\section{Conclusion}

In summary, we developed an extension of the split-operator technique which allows for the study of systems with spin-dependent Hamiltonians. The advantage of this technique lies in the fact that it is easy to implement and it allows for separating real and reciprocal parts of the time evolution operator, so that one avoids writing the momentum in terms of derivatives. We exemplify the use of this technique in two cases of great current interest: (i) the Zeeman and spin-orbit effects in semiconductor quantum dots and 2DEG, and (ii) the Klein tunneling and trembling motion of wave packets in graphene.

\section*{Acknowledgments}
The authors gratefully acknowledge fruitful discussions with J. M. Pereira Jr. and R. N. Costa Filho. This work was financially supported by CNPq through the INCT-NanoBioSimes and the Science Without Borders programs (contract 402955/2012-9), PRONEX/FUNCAP, CAPES, the Bilateral programme between Flanders and Brazil, and the Flemish Science Foundation (FWO-Vl).


\end{document}